\documentclass[doublecol]{epl2} 
\usepackage[utf8]{inputenc}

\title{Multifractal nature of seismic sequences distributed along the Pacific Ring of Fire}
\shorttitle{Multifractal nature along the Pacific Ring of Fire} 

\author{D. B. de Freitas\inst{1} \and G. S. Fran\c{c}a\inst{2}}
\shortauthor{de Freitas \etal}

\institute{                    
  \inst{1} Departamento de F\'{\i}sica, Universidade Federal do Cear\'a, Caixa Postal 6030, Campus do Pici, 60455-900 Fortaleza, Cear\'a, Brazil\\
  \inst{2} Observat\'orio Sismol\'ogico-IG/UnB, Campus Universit\'ario Darcy Ribeiro SG 13 Asa Norte, 70910-900 Bras\'{\i}lia, Brazil
}
\pacs{nn.mm.xx}{First pacs description}
\pacs{nn.mm.xx}{Second pacs description}
\pacs{nn.mm.xx}{Third pacs description}

\abstract{
A multifractal methodology was utilized to analyze a set of seismic sequences distributed along the Pacific Ring of Fire, sourced from the National Earthquake Information Center (NEIC) catalog. The analysis employed the Multifractal Detrended Moving Average Analysis (MFDMA) method to characterize the multi-scale behavior using various geometrical parameters of the multifractal spectrum. The findings of this study can be summarized as follows: firstly, our research suggests that seismic sequences along the Ring of Fire exhibit distinct dynamics; and secondly, it indicates that long-range correlations may influence larger magnitude earthquakes, as demonstrated by the correlation with the $b_{\rm GR}$ index. These results contribute to an enhanced understanding of the multifractal characteristics of seismic activity and their implications for earthquake dynamics.}

\begin{document}

\maketitle

\section{Introduction}

Seismic sequences along Circum-Pacific subduction zones have been the subject of extensive research. de Freitas \textit{et al}. \cite{defreitas2019} analyzed the signatures of long-range persistence in seismic sequences along these zones, from Chile to Kermadec, using data from the National Earthquake Information Center (NEIC) catalog. Lay \& Kanamori \cite{lay} interpreted the variation in maximum rupture extent of large shallow earthquakes in circum-Pacific subduction zones in the context of the asperity model of stress distribution on the fault plane. Additionally, Tsapanos \cite{tsa} examined the temporal behavior of aftershock sequences distributed on the subduction zones in the circum-Pacific belt. Moreover, Schwartz \& Rokosky \cite{sch} presented a review of slow slip events and related seismic tremors observed at plate boundaries worldwide, with a focus on circum‐Pacific subduction zones. These studies provide valuable insights into the behavior and characteristics of seismic sequences along Circum-Pacific subduction zones.

In the majority of cases, geophysical signals exhibit irregular and complex temporal fluctuations, characterized by inhomogeneous variations and extreme events, such as irregular rupture propagation and non-uniform distributions of rupture velocity, stress drop, and co-seismic slip \cite{telesca0}. The presence of scaling properties in geophysical data suggests that the fractal method may offer a viable approach to investigating the behavior of earthquake magnitude fluctuations \cite{li}. On one hand, terrestrial tectonic activity arises from highly complex mechanisms involving numerous variables, including deformation, rupture, released energy, land features, and heterogeneity in the seismogenic plate interface \cite{kawa,scherrer}. On the other hand, a variety of methods and tools are available for a more comprehensive description of the dynamic properties of earthquakes \cite{omori,gr}.

In particular, several statistical methods are available in the scientific literature, which use the concept of (multi)fractality. Among them, we can find methods based on self-similar and self-affine fractals such as the box dimension \cite{peitgen2004chaos}, the detrended fluctuation analysis (DFA) \cite{1992Natur.356..168P}, the detrending moving average analysis (DMA) \cite{ale}, the scaled windowed variance analysis (SWVA) \cite{1985PhyS...32..257M}, and so on. In general, multifractal analysis and its different methods and procedures \cite{Kantelhardt,gu2010,tang}, which were developed over more than five decades, are applied in many varied fields of knowledge as inspired by \cite{hurst1951,mw1969a,mw1969b,feder1988}. In several areas, such as medicine \cite{ivanov1999} and geophysics \cite{telesca2006,defreitas2013}, multifractality has already been adopted as a determinant approach for analyzing the behaviors of time series with nonlinearity, nonstationarity and correlated noise, which are just a few of the properties that this analysis can describe \cite{movahed,Norouzzadeha,sps2009,seuront,a2011}.

In this context, we decided to focus our attention on the seminal parameter proposed by Hurst \cite{hurst1951} to describe the long-term dependence of water levels in rivers and reservoirs \cite{seuront}. Unlike the current trend that directly applies the multifractal methods \cite{telesca1,telesca2,telesca2006}, we decided to investigate the dynamics of earthquakes using a set of five multifractal indicators, where features as memory and long-term correlations are investigated by surrogates (shuffled and randomized data) of time series. As quoted by Telesca \cite{telesca2} and Telesca \textit{et al}. \cite{telesca2006}, a statistical analysis based on multifractals is featured by power-laws and can be a powerful tool to examine the temporal fluctuations at different scales when applied to earthquake magnitude time series.

In the present paper, we investigate the long-term persistence signatures present in seismic sequences along Circum-Pacific subduction zones initially treated by Scheerer \textit{et al}. \cite{scherrer}. In this sense, our study applied the detrending moving average algorithm for one-dimensional multifractal signals (MFDMA) to seismic data. It is worth noting the universal character of the $R/S$ method in the analysis of the behavior of fluctuations. A large number of studies at different areas of knowledge has shown that the so-called Hurst exponent extracted from within the MFDMA analysis provides a robust and powerful statistical method to characterize nonstationary fluctuations at different timescales \cite{sps2009,defreitas2019}. In a previous paper using the classical fractal $R/S$ method, \cite{defreitas2013a} found that for San Andreas fault the Hurst exponent of 0.87, indicated a strong long-term persistence. Other studies (e.g., \cite{li}) also indicate that the Hurst exponent of seismic data calculated by $R/S$ method is greater than 0.6.

Our main interest is to examine a possible correlation between the scaling properties of the subduction-zone earthquakes on the Circum-Pacific controlled by the interaction of asperities \cite{lay} and different multifractal indicators estimated from the MFDMA analysis \cite{mw1969a,gu2010}. In general, we believe that different subduction zones distributed in major groups can be associated with distinct scaling laws that relate the dynamics of the earthquakes and their short- and long-term fluctuations. Moreover, this procedure could be used to distinguish the zones with the distribution of stronger stress from the weaker ones. As mentioned by Lay and Kanamori \cite{lay}, the interaction and failure of an asperity can cause an increase in stress on the adjacent asperities. In addition, they described the main properties of subduction zones in Table 2 from the referred paper. Indeed, our theoretical background is based on this idea of asperity proposed by these authors. Finally, the authors elaborate a general structure of categories based on the extreme behavior of dynamics and strength of the earthquakes as cited in Scheerer \textit{et al}. \cite{scherrer}.

Our paper is organized as follows: in the next section, we describe the multifractal method used in our study and a detailed discussion about the set of five multifractal indicators extracted from the multifractal spectrum. In Section 3, we present our seismic sample. The main results and their physical implications are presented in section 4, and the final remarks are highlighted in the last section.

\section{The multifractal background}
Since 2016, de Freitas \textit{et al}. \cite{defreitas2016,defreitas2017,defreitas2019} used multifractal analysis in astrophysical data that were observed by the Kepler mission, as well as the Sun in its active phase. de Freitas \textit{et al}. \cite{defreitas2016,defreitas2017} showed that the multifractal detrending moving average (MFDMA) algorithm, which was developed by \cite{gu2010}\footnote{MATLAB codes for MFDMA analysis can be found in the \texttt{arXiv} version of \cite{gu2010}'s paper: \texttt{https://arxiv.org/pdf/1005.0877v2.pdf}} and \cite{tang}, is a powerful technique that provides invaluable information on the dynamic structure of a time series.

We summarized the MFDMA algorithm in the following steps according to \cite{gu2010}:

\begin{itemize}
	\item Step 1: Calculate the time-series profile over time $t=1,2,3,...,N$:
\end{itemize}
\begin{equation}
\label{eq1}
y(t)=\sum^{t}_{i=1}x(i), \quad t=1,2,3,...,N,
\end{equation}
where the above equation is a sequence of cumulative sums. As introduced by \cite{defreitas2016,defreitas2017}, we adopt $\theta$=0 in the equations below.

\begin{itemize}
	\item Step 2: Calculate the moving average function of Eq. (\ref{eq1}) in a moving window:
\end{itemize}
\begin{equation}
\label{eq1a}
\tilde{y}(t)=\frac{1}{s}\sum^{\left\lceil s-1\right\rceil}_{k=0}y(t-k),
\end{equation}
where $s$ is the window size, and $\left\lceil (x)\right\rceil$ is the smallest integer that is not smaller than argument $(x)$.

\begin{itemize}
	\item Step 3: Detrend the series by removing the moving average function, $\tilde{y}(i)$, and obtain the residual sequence, $\epsilon(i)$, through:
\end{itemize}
\begin{equation}
\label{eq2}
\epsilon(i)=y(i)-\tilde{y}(i).
\end{equation}
where $s \le i\le N$. The residual series, $\epsilon(i)$, is divided into $N_{s}$ disjoint segments with the same size of $s$, where $N_{s}=\left\lfloor N/s -1\right\rfloor$. In addition, each segment can be expressed by $\epsilon_{\nu}$, where $\epsilon_{\nu}(i)=\epsilon(l+i)$ such that $1\le i \le s$ and $l=(\nu-1)s$.

\begin{itemize}
	\item Step 4: Calculate the root-mean-square (RMS) fluctuation function, $F_{\nu}(s)$, for a segment of size $s$:
\end{itemize}
\begin{equation}
\label{eq3}
F_{\nu}(s)=\left\{\frac{1}{s}\sum^{s}_{i=1}\epsilon^{2}_{\nu}(i)\right\}^{\frac{1}{2}}.
\end{equation}

\begin{itemize}
	\item Step 5: Generate the function, $F_{q}(s)$, of the $q$th order:
\end{itemize}
\textbf{\begin{equation}
	\label{eq4}
	F_{q}(s)=\left\{\frac{1}{N_{s}}\sum^{N_{s}}_{\nu=1}F^{q}_{\nu}(s)\right\}^{\frac{1}{q}}, 
	\end{equation}}
for all $q\neq 0$, where the $q$th-order function is the statistical moment (\textit{e.g.}, for $q$=2, we have the variance), and for $q=0$,

\textbf{\begin{equation}
	\label{eq4b}
	\ln\left[F_{0}(s)\right]=\frac{1}{N_{s}}\sum^{N_{s}}_{\nu=1}\ln [F_{\nu}(s)], 
	\end{equation}}
where the scaling behaviour of $F_{q}(s)$ follows the relationship that is given by $F_{q}(s)\sim s^{h(q)}$, and $h(q)$ denotes the Holder exponent or generalized Hurst exponent. Each value of $q$ yields a slope $h$.

\begin{itemize}
	\item Step 6: Knowing $h(q)$, the multifractal scaling exponent, $\tau(q)$, can be computed:
\end{itemize}
\begin{equation}
\label{eq5}
\tau(q)=q h(q)-1.
\end{equation}	

Finally, the singularity strength function, $\alpha(q)$, and the multifractal spectrum, $f(\alpha)$, are obtained via a Legendre transform:
\begin{equation}
\label{eq7}
\alpha(q)=\frac{d\tau(q)}{dq}
\end{equation}	
and
\begin{equation}
\label{eq6}
f(\alpha)=q\alpha-\tau(q).
\end{equation}	
In addition, for a monofractal signal, $h$ is the same for all values of $q$. For a multifractal signal, $h(q)$ is a function of $q$, and the multifractal spectrum is parabolic (see \cite{defreitas2017}, Fig. 2). In particular, for $q=2$, we have a specific value of $h$ denoted by $H$, where $H$ is the global Hurst exponent.

We use the following model parameters to yield the multifractal spectrum, as recommended by \cite{gu2010}: $N$=30; $q\in[-5,5]$ with a step size of 0.2; $\theta$=0; the lower bound of segment size $s$, which is denoted as $s_{min}$ and set to 10; and the upper bound of segment size $s$, which is denoted as $s_{max}$ and is given by \textbf{$N/10$}.

\subsection{Multifractal indicators}
We tested the set of four multifractal descriptors that were extracted from the spectrum $f(\alpha)$, as proposed by de Freitas \textit{et al.} \cite{defreitas2017}. An illustration of indicators used to quantify the multifractal spectrum in this work is shown in Figure 2 from \cite{defreitas2017}. In this paper, the authors have used this same figure to describe the shape of the multifractal spectrum. Here, it is listed five indicators, as mentioned by \cite{defreitas2017}:

1) The parameter $\alpha_{0}$: The $\alpha_{0}$ parameter delivers valuable information about the structure of the studied process, with a high value indicating that it is less correlated and processes fine structure \cite{krzyszczak2019}. In addition, this parameter is strongly affected by signal variability. This will be evidenced when we investigate the different sources of multifractality that are present in the seismic signal. 

2) Degree of asymmetry ($A$):
This index, which also called the skewness in the shape of the $f(\alpha)$ spectrum, is expressed as the following ratio:
\begin{equation}
\label{eq8}
A=\frac{\alpha_{\rm max}-\alpha_{0}}{\alpha_{0}-\alpha_{\rm min}},
\end{equation}
where $\alpha_{0}$ is the value of $\alpha$ when \textbf{$f(\alpha)$} is maximal. The value of this index $A$ indicates one of three shapes: right-skewed ($A>1$), left-skewed ($0<A<1$) or symmetric ($A=1$). The left endpoint $\alpha_{\rm min}$ and the right endpoint $\alpha_{\rm max}$ represent the maximum and minimum values of the singularity exponent, respectively.

3) Degree of multifractality ($\Delta \alpha$):
This index represents the broadness:
\begin{equation}
\label{eq9}
\Delta \alpha=\alpha_{\rm max}-\alpha_{\rm min},
\end{equation} 
where $\alpha_{\rm max}$ and \textbf{$\alpha_{\rm min}$} are as defined above. A low value of $\Delta\alpha$ indicates that the time series is close to fractal, and the multifractal strength is higher when $\Delta\alpha$ increases \cite{defreitas2009,defreitas2017}.

4) Singularity parameter $\Delta f_{\rm min}(\alpha)$:
Parameter $\Delta f_{\rm min}(\alpha)$ characterizes the broadness, which is defined as the difference $f(\alpha_{\rm max})-f(\alpha_{\rm min})$ of the singularity spectrum. If $\Delta f_{\rm min}(\alpha)>1$, the left-hand side is less deep, while if $C<1$, this side is deeper, and if $\Delta f_{\rm min}(\alpha)=1$, the depths of the tails are the same on both sides. As quoted by Ihlen \cite{ihlen}, a long left tail implies that the singularities are stronger, whereas a long right tail indicates that the singularities are weaker \cite{tanna}. 

5) The global Hurst exponent ($H$): In the multifractal context, the Hurst exponent $H$ is defined by the second-order statistical moment (i.e., variance or standard deviation) using eq. (\ref{eq5}), which is denoted by $q=2$ \cite{hurst1951,ihlen}. A broad explanation of the exponent $H$ can be found in \cite{defreitas2017}. 	
According to \cite{defreitas2013}, the exponent $H$ denotes Brownian motion when $H=1/2$, i.e., past and future fluctuations are uncorrelated. However, if $H>1/2$, fluctuations are linked to the long-term persistence signature, i.e., an increase in values will most likely be followed by another increase in the short term, and a decrease in values will most likely be followed by another decrease in the short term. Already for $H<1/2$, the fluctuations tend not to continue in the same direction, but instead turn back on themselves, which results in a less-smooth time series \cite{hm}. 

\subsection{Origin of multifractality}

\begin{figure}
\begin{center}
	\includegraphics[width=0.99\columnwidth]{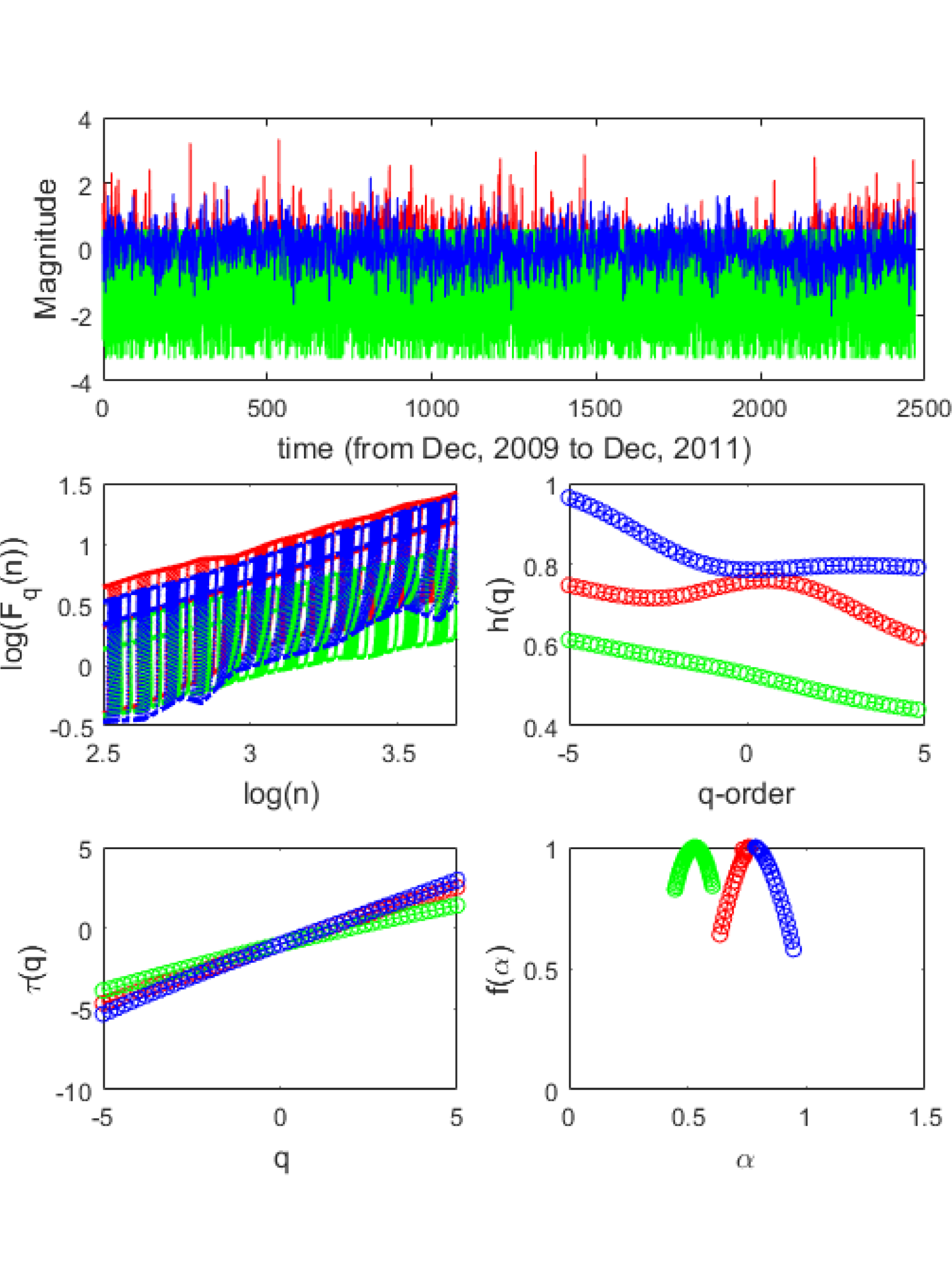}
	\end{center}
	\caption{Multifractal analysis of magnitude for Aleutians Zone following steps 5 and 6 present in Section 2. \textit{Top panel}: The original (in red), the shuffled (green) and phase randomized surrogate (blue) data are based on the procedure mentioned in Section 2 \textit{Left middle}: the multifractal fluctuation function $F_{q}(n)$ obtained from MFDMA method. Each curve corresponds to different fixed values of $q=-5,...,5$ (with a step of 0.2) from top to bottom, where red lines are original data, green lines are shuffled data and blue lines are surrogate data. \textit{Right middle}: $q$-order Hurst exponent ($h(q)$) as a function of $q$-parameter. This panel shows the truncation originated from the leveling of the $h(q)$ for positive $q$'s. \textit{Left bottom}: comparison of the multifractal scaling exponent $\tau(q)$ of the original (red), shuffled (green) and surrogate (blue) data. In this panel is possible to identify a crossover in $q=0$, a typical feature of the multifractal signal. \textit{Right bottom}: multifractal spectrum $f(\alpha)$ of the original (red), shuffled (green) and surrogate (blue) time series.}
	\label{fig0}
\end{figure}

Figure \ref{fig0} shows two other types of time series, namely, shuffling (green circles) and phase randomized (blue circles) data, which are used to verify the different origins of the multifractality. In all earthquake magnitude time series of our sample, the analysis emphasized here indicates that the two types source of multifractality occur.

In particular, shuffling series destroys the memory but preserves the distribution of the data with $h(q)=0.5$. In this case, the source of the multifractality in the time series only presents long-range correlations \cite{defreitas2017}. However, the origin of multifractality can also be due to the presence of non-linearity in the original time series (red circles in Fig.~\ref{fig0}. General speaking, the non-linear effects can be weakened by creating phase-randomized surrogates, thereby preserving the amplitudes of the Fourier transform and the linear properties of the original series by randomizing the Fourier phases \cite{Norouzzadeha,defreitas2017}. In this case, if the origin of multifractality is non-linearity that is obtained by the phase randomized method, the values of $h(q)$ will be independent of $q$, and $h(q)=0.5$ will not necessarily hold.

These sources can be checked by using generalized Hurst exponent $h(q)$ from original series with result from corresponding shuffled $h_{\rm shuf} (q)$ and phase randomized $h_{\rm pran} (q)$ surrogates of time series. Differences between these two $h'$s with original one indicate the presence of long-range correlations or nonlinearity in the original time series, respectively.

The $h(q)$ obtained from MDFMA is related to the Renyi exponent $\tau(q)$ by relationship:
$\tau(q)=qh(q)-1$,
Where the differences
$\Delta\tau_{\rm shuf} (q)=h(q)-h_{\rm shuf} (q)$, and
$\Delta\tau_{\rm pran} (q)=h(q)-h_{\rm pran} (q)$, directly indicate the presence of long-range correlations or nonlinearity in the original time series.

If only nonlinearity is responsible for the multifractality, one should find $h(q)=h_{\rm shuf} (q)$ and, therefore, $\Delta\tau_{\rm shuf} (q)=0$. On the other hand, deviations from $\Delta\tau_{\rm shuf} (q)=0$ indicates the presence of correlations, and $q$ dependence of $\Delta\tau_{\rm shuf} (q)$ indicates that multifractality is due to the long-range correlation. If only correlation multifractality is present, one finds $h_{\rm shuf} (q)=0.5$. If both nonlinearity and correlation are present, both $h_{\rm shuf} (q)$ and $h_{\rm pran} (q)$ are a function of $q$. Absolute values of $\Delta\tau_{\rm pran} (q)$ is greater than $\Delta\tau_{\rm shuf} (q)$, so multifractality due to correlation is weaker than multifractality due to nonlinearity. We used the following criteria to define the source of multifractality: ``0'', if the multifractality due to correlation is weaker than that due to nonlinearity, and ``1'', if the multifractality due to correlation is stronger than that due to nonlinearity. These criteria are applied in all the sample and result is summarized in Table \ref{tab1}.


\begin{figure}
\begin{center}
\includegraphics[width=0.49\textwidth]{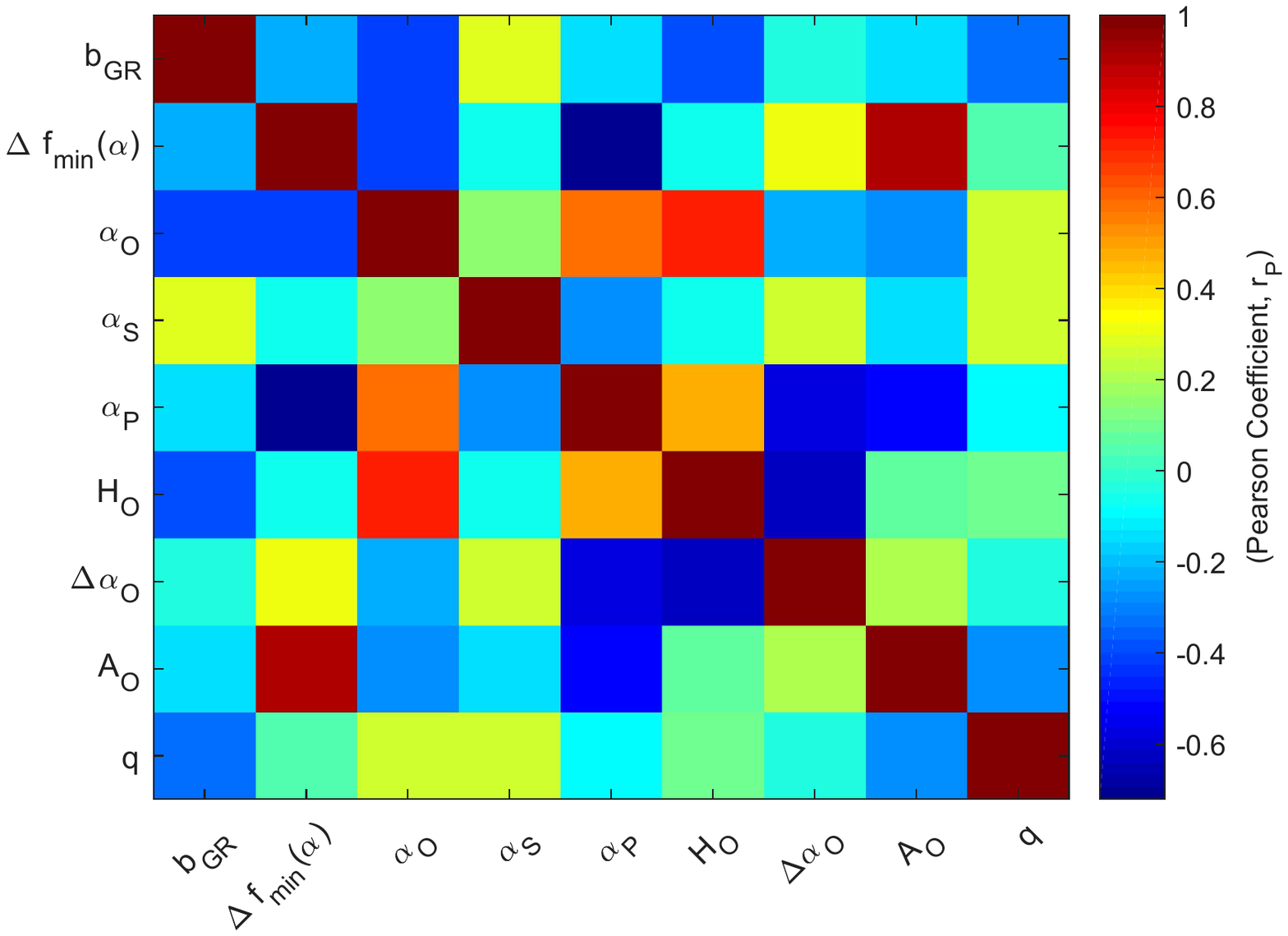}
\end{center}
\caption{Matrix of Pearson correlation coefficients showing the simple linear relationship among all the parameters extracted from our sample.}
\label{fig1}
\end{figure}

\section{Working sample and data analysis}
Recently de Freitas \textit{et al}. \cite{defreitas2019} published a study about the fractal properties of a set of circum-Pacific subduction zones distributed along the so-called Ring of Fire. As mentioned in referred paper, these data were extracted from the National Earthquake Information Center (NEIC) catalog and they were selected 12 regions Flinn-Engdhal for applying our analysis as shown in Table \ref{tab1} \cite{young}. As reported by Scherrer \textit{et al}. \cite{scherrer} and \cite{defreitas2019}, the NEIC catalog offers magnitude time series in different magnitudes types ($M_{\rm w},M_{\rm b},M_{\rm s},M_{\rm l},M_{\rm d}$) for the same event and, therefore, we choose to follow NEIC automatic ranking. In addition, we consider that using this sequence makes no significant impact on the final result of the present paper because the differences between magnitudes types are small. 

According to Scherrer \textit{et al}. \cite{scherrer} the data sample is distributed in four different subduction zones defined by asperity and broadness of rupture front. The main structure of zones is described by these authors and can be found in Figure 1. In this context, the reader is referred to Scherrer \textit{et al}. \cite{scherrer} for instrumental procedure and classification. For the present analysis, we considered a magnitude greater than 3, in this case, the only effect from macroearthquakes is analyzed. We also compared the five multifractal indicators and the classical index $b_{\rm GR}$ from Gutenberg-Richter law \cite{gr} by ZMAP software measured by Scherrer \textit{et al}. \cite{scherrer}. 

\begin{table}
	\caption{Identifier number of $SZ$ and the number that indicates the type of multifractality for $q<0$ and $q>0$.}
	\label{tab1}
	\begin{center}
		\begin{tabular}{lccccc}
			\hline
			Area & $SZ$ & $q<0$  & $q>0$ \\
			&  &  & & \\
			\hline
			Alaska & 1& 0 & 1\\
			Aleutians & 2& 1 & 1\\
			Central America & 2-3& 0 & 0\\
			Central Chile & 3 & 1 & 1\\
			Colombia & 2 & 0 & 1\\
			Kuriles & 3 & 0 & 1\\
			Marianas & 4 & 0 & 1\\
			New Hebrides & 2-3& 0 & 1\\
			Peru &3 & 0 & 0\\
			Solomon Islands &2 & 0 & 1\\
			Tonga \& Kermadec & 4 & 0 & 1\\
			\hline
		\end{tabular}
	\end{center}
\end{table}

\section{Results and discussions}
As shown in the top panels from Fig. \ref{fig0}, the MFDMA method was used to estimate the values of five multifractal parameters for a data set of 12 circum-Pacific subduction zones. As a result, the values of these indicators were calculated using the multifractal spectrum for each zone as indicated in the right bottom panel from Fig. \ref{fig0}.

In order to reduce the number of figures, we made the decision to display only the results from the Aleutians Zone (see Fig. \ref{fig0}). Generally speaking, the results are similar across all subduction zones, with a longer tail on the right side of the multifractal spectrum observed in the original time series (depicted in red). It is important to highlight that the surrogate data exhibits similarity across the different zones. When it comes to shuffled data, the $\alpha$-value for all zones decreased to 0.5, indicating a strong presence of long-range correlation or memory. Conversely, the randomized phase data for all zones show a significant deviation from the original data, suggesting the existence of non-linear patterns in the time series. This outcome provides sufficient evidence to infer that distinct physical processes govern the dynamics of abduction zones. Upon observing the figure illustrating the behavior of $h(q)$ versus $q$, it becomes evident that for $q$ greater than zero (in modulus), the phase randomized surrogate data (depicted in blue) differs from the values for the original time series.

Following the methodology outlined in Section 2, a procedure has been established to determine the dominant source of multifractality. This involves the segregation of multifractality sources as a function of $q$. As indicated in Table \ref{tab1}, a clear differentiation between multifractality sources is evident when $q$ is divided into the intervals $q>0$ (strong singularity) and $q<0$ (weak singularity). Primarily, the source associated with long-range correlation prevails at $q>0$, whereas this pattern is reversed at $q<0$, with the regime of nonlinearity assuming prominence. Notably, there are only four abduction zones where only one source of multifractality predominates for both $q$ ranges, denoted as 1-1 or 0-0. This behavior is particularly prominent in zones 2 and 3, with zone 3 exhibiting a greater emphasis. Fundamentally, these zones demonstrate the coupling of tectonic plates between zones 1 and 4, representing the extremes in earthquake rupture length, with 1 denoting the strongest and 4 representing the weakest in coupling \cite{scherrer}. However, the subtle advantage of zone 3 over zone 2 may potentially be attributed to the heterogeneity of stress distribution in Kuriles-type zones, which are characterized by smaller ruptures but appear to exhibit significant mechanical coupling, resulting in a more pronounced formation and consumption of fragments when considering the entire contact area.

Figure \ref{fig1} provides a comprehensive overview of the behavior of multifractal indicators in comparison with the classical index $b_{\rm GR}$ derived from the Gutenberg-Richter law. The figure presents Pearson coefficients, considering a linear correlation between the parameters as the null hypothesis. It is noteworthy that the subscripts O, S, and P denote the original series and the surrogate data, respectively. Additionally, it is important to emphasize that only the parameter $\alpha$ was considered for the different types of time series, as it serves as a crucial marker for the position of the multifractal spectrum, enabling the distinction between shuffled data and phase-randomized data.

As depicted in Figure \ref{fig1}, all multifractal indicators exhibit anti-correlation with the $b_{\rm GR}$ index, with the highest value observed for $\alpha_{0}$ and $H_{0}$. Furthermore, a strong anti-correlation is evident with the value of $q$ (0 or 1). Specifically, the decrease in bGR values is associated with fluctuations in the time series attributed to strong long-range correlations. Similarly, for the indicators $\alpha_{0}$ and $H_{0}$, the anti-correlation arises from the fact that as $b_{\rm GR}$ decreases, the respective values of these parameters enter the signal persistence regime, signifying that the fluctuations become less abrupt. Apart from the correlation with the $b_{\rm GR}$ index, the majority of the other correlations also demonstrate a strong anti-correlation.

In conclusion, while there are numerous interpretations of the results depicted in the aforementioned figure regarding the correlations between the multifractal indicators, it is impractical to enumerate each one. Nevertheless, the identified anti-correlations yield substantial information regarding how fluctuations with weak and strong singularity impact the width and depth of the right tail of the multifractal spectrum. This is crucial for understanding that the increase in earthquake intensity may be influenced by the contribution of a specific source of multifractality. Notably, evidence suggests that long-range correlations may be accountable for larger magnitude earthquakes, as lower values of $b_{\rm GR}$ correspond to a longer tail in the Gutenberg-Richter law.

\section{Final remarks}
The study utilized the MFDMA method to estimate multifractal parameters for a dataset of 12 circum-Pacific subduction zones. The results indicated similar behavior across all subduction zones, with distinct physical processes governing the dynamics of abduction zones. The multifractal indicators exhibited anti-correlation with the classical index $b_{\rm GR}$, providing insights into the impact of weak and strong singularity fluctuations on earthquake intensity. The dominant source of multifractality was determined, revealing the prevalence of long-range correlation and nonlinearity in different $q$ ranges. The subtle advantage of certain zones was attributed to the heterogeneity of stress distribution. Overall, the findings suggest that long-range correlations may influence larger magnitude earthquakes, as evidenced by the correlation with the $b_{\rm GR}$ index. These results contribute to a better understanding of the multifractal characteristics of seismic activity and their implications for earthquake dynamics.

\acknowledgments
DBdeF acknowledges financial support from the Brazilian agency CNPq-PQ2 (Grant No. 305566/2021-0). Research activities of STELLAR TEAM of Federal University of Cear\'a are supported by continuous grants from the Brazilian agency CNPq.


\bibliographystyle{eplbib}
\bibliography{MDFMAFireRing}

\begin{thebibliography}{10}
\expandafter\ifx\csname url\endcsname\relax\def\url#1{\texttt{#1}}\fi

\bibitem{defreitas2019}
\Name{de~Freitas D.~B., França G., Scherrer T., Vilar C. \and Silva R.} \REVIEW{Brazilian Journal of Geophysics}{37}{2019}{409}.

\bibitem{lay}
\Name{Lay T. \and Kanamori H.} \Book{An Asperity Model of Large Earthquake Sequences} (American Geophysical Union (AGU)) 1981 Ch.~2 pp. 579--592.

\bibitem{tsa}
\Name{Tsapanos T.} \REVIEW{Geophysical Journal International}{123}{1995}{633}.

\bibitem{sch}
\Name{Schwartz S. \and Rokosky J.} \REVIEW{Reviews of Geophysics}{45}{2007}{}.

\bibitem{telesca0}
\Name{{Telesca} L., {Colangelo} G. \and {Lapenna} V.} \REVIEW{Natural Hazards and Earth System Sciences}{5}{2005}{673}.

\bibitem{li}
\Name{{Li} J. \and {Chen} Y.} \REVIEW{Acta Seismologica Sinica}{14}{2001}{148}.

\bibitem{kawa}
\Name{{Kawamura} H., {Hatano} T., {Kato} N., {Biswas} S. \and {Chakrabarti} B.~K.} \REVIEW{Reviews of Modern Physics}{84}{2012}{839}.

\bibitem{scherrer}
\Name{{Scherrer} T.~M., {Fran{\c{c}}a} G.~S., {Silva} R., {de Freitas} D.~B. \and {Vilar} C.~S.} \REVIEW{Physica A Statistical Mechanics and its Applications}{426}{2015}{63}.

\bibitem{omori}
\Name{Omori F.} \REVIEW{The journal of the College of Science, Imperial University, J.}{7}{1895}{111}.

\bibitem{gr}
\Name{{Gutenberg} B. \and {Richter} C.~F.} \REVIEW{The Bulletin of the Seismological Society of America}{34}{1944}{185}.

\bibitem{peitgen2004chaos}
\Name{Peitgen H., J{\"u}rgens H. \and Saupe D.} \Book{Chaos and Fractals: New Frontiers of Science} (Springer New York) 2004.

\bibitem{1992Natur.356..168P}
\Name{{Peng} C.~K., {Buldyrev} S.~V., {Goldberger} A.~L., {Havlin} S., {Sciortino} F., {Simons} M. \and {Stanley} H.~E.} \REVIEW{Nature}{356}{1992}{168}.

\bibitem{ale}
\Name{{Alessio, E.}, {Carbone, A.}, {Castelli, G.} \and {Frappietro, V.}} \REVIEW{Eur. Phys. J. B}{27}{2002}{197}.

\bibitem{1985PhyS...32..257M}
\Name{{Mandelbrot} B.~B.} \REVIEW{Physica Scripta}{32}{1985}{257}.

\bibitem{Kantelhardt}
\Name{Kantelhardt J., Zschiegner S., Koscielny-Bunde E., Havlin S., Bunde A. \and Stanley H.} \REVIEW{Physica A: Statistical Mechanics and its Applications}{316}{2002}{87}.

\bibitem{gu2010}
\Name{{Gu} G.-F. \and {Zhou} W.-X.} \REVIEW{Physical Review E}{82}{2010}{011136}.

\bibitem{tang}
\Name{Tang L., Lv H., Yang F. \and Yu L.} \REVIEW{Chaos, Solitons \& Fractals}{81}{2015}{117}.

\bibitem{hurst1951}
\Name{Hurst H.~E.} \REVIEW{Transactions of the American Society of Civil Engineers}{116}{1951}{770–799}.

\bibitem{mw1969a}
\Name{Mandelbrot B.~B. \and Wallis J.~R.} \REVIEW{Water Resources Research}{5}{1969}{321–340}.

\bibitem{mw1969b}
\Name{Mandelbrot B.~B. \and Wallis J.~R.} \REVIEW{Water Resources Research}{5}{1969}{967–988}.

\bibitem{feder1988}
\Name{Feder J.} \Book{Fractals} (Plenum Press, New York) 1988.

\bibitem{ivanov1999}
\Name{{Ivanov} P.~C., {Amaral} L. A.~N., {Goldberger} A.~L., {Havlin} S., {Rosenblum} M.~G., {Struzik} Z.~R. \and {Stanley} H.~E.} \REVIEW{Nature}{399}{1999}{461}.

\bibitem{telesca2006}
\Name{Telesca L., Lapenna V. \and Macchiato M.} \REVIEW{Geological Society, London, Special Publications}{261}{2006}{95}.

\bibitem{defreitas2013}
\Name{de~Freitas D.~B., Leão I.~C., Lopes C. E.~F., Paz-Chinchon F., Martins B. L.~C., Alves S., Medeiros J. R.~D. \and Catelan M.} \REVIEW{The Astrophysical Journal Letters}{773}{2013}{L18}.

\bibitem{movahed}
\Name{Movahed M.~S., Jafari G.~R., Ghasemi F., Rahvar S. \and Tabar M. R.~R.} \REVIEW{Journal of Statistical Mechanics: Theory and Experiment}{2006}{2006}{P02003}.

\bibitem{Norouzzadeha}
\Name{{Norouzzadeh} P., {Dullaert} W. \and {Rahmani} B.} \REVIEW{Physica A Statistical Mechanics and its Applications}{380}{2007}{333}.

\bibitem{sps2009}
\Name{{Suyal} V., {Prasad} A. \and {Singh} H.~P.} \REVIEW{Solar Physics}{260}{2009}{441}.

\bibitem{seuront}
\Name{Seuront L.} \Book{Fractals and multifractals in ecology and aquatic science} (CRC Press, United Kingdom) 2010.

\bibitem{a2011}
\Name{{Aschwanden} M.~J.} \Book{{Self-Organized Criticality in Astrophysics}} 2011.

\bibitem{telesca1}
\Name{{Telesca} L. \and {Lapenna} V.} \REVIEW{Tectonophysics}{423}{2006}{115}.

\bibitem{telesca2}
\Name{Telesca L.} \Book{Fractal Methods in the Investigation of the Time Dynamics of Fires: An Overview} (Springer International Publishing, Cham) 2016 pp. 117--152.

\bibitem{defreitas2013a}
\Name{de~Freitas D.~B., França G.~S., Scherrer T.~M., Vilar C.~S. \and Silva R.} \REVIEW{Europhysics Letters}{102}{2013}{39001}.

\bibitem{defreitas2016}
\Name{de~Freitas D.~B., Nepomuceno M. M.~F., de~Moraes~Junior P. R.~V., Lopes C. E.~F., Chagas M. L.~D., Bravo J.~P., Costa A.~D., Martins B. L.~C., Medeiros J. R.~D. \and Leão I.~C.} \REVIEW{The Astrophysical Journal}{831}{2016}{87}.

\bibitem{defreitas2017}
\Name{de~Freitas D.~B., Nepomuceno M. M.~F., de~Souza M.~G., Leão I.~C., Chagas M. L.~D., Costa A.~D., Martins B. L.~C. \and Medeiros J. R.~D.} \REVIEW{The Astrophysical Journal}{843}{2017}{103}.

\bibitem{krzyszczak2019}
\Name{{Krzyszczak} J., {Baranowski} P., {Zubik} M., {Kazandjiev} V., {Georgieva} V., {S{\l}awi{\'n}ski} C., {Siwek} K., {Kozyra} J. \and {Nier{\'o}bca} A.} \REVIEW{Theoretical and Applied Climatology}{137}{2019}{1811}.

\bibitem{defreitas2009}
\Name{de~Freitas D.~B. \and Medeiros J. R.~D.} \REVIEW{Europhysics Letters}{88}{2009}{19001}.

\bibitem{ihlen}
\Name{Ihlen E.} \REVIEW{Frontiers in Physiology}{3}{2012}{}.

\bibitem{tanna}
\Name{{Tanna} H.~J. \and {Pathak} K.~N.} \REVIEW{Astrophysical Supplementary Seriesa2011}{350}{2014}{47}.

\bibitem{hm}
\Name{Hampson K.~M. \and Mallen E. A.~H.} \REVIEW{Biomed. Opt. Express}{2}{2011}{464}.

\bibitem{young}
\Name{Young J., Presgrave B., Aichele H., Wiens D. \and Flinn E.} \REVIEW{Physics of the Earth and Planetary Interiors}{96}{1996}{223}.

\end{thebibliography}

\end{document}